\documentclass[aps,pre,twocolumn,nofootinbib,floatfix,showpacs]{revtex4}  

\usepackage{graphicx} 
\usepackage{amsmath}
\newcommand {\be}{\begin{equation}}
\newcommand {\ee}{\end{equation}}
\newcommand {\ba}{\begin{eqnarray}}
\newcommand {\ea}{\end{eqnarray}}

\begin{document}

\title{Structure of characteristic Lyapunov vectors in anharmonic 
Hamiltonian lattices}

\author{M. Romero-Bastida}
\email{mromerob@ipn.mx}
\affiliation{SEPI-ESIME Culhuac\'an, Instituto Polit\'ecnico Nacional, Av. Santa
Ana No. 1000, Col. San Francisco Culhuac\'an, Delegaci\'on Coyoacan, Distrito
Federal 04430, M\'exico}
\author{Diego Paz\'o}
\email{pazo@ifca.unican.es}
\affiliation{Instituto de F\'\i sica de Cantabria (IFCA), CSIC-Universidad de
Cantabria, E-39005 Santander, Spain}
\author{Juan M. L\'opez}
\email{lopez@ifca.unican.es}
\affiliation{Instituto de F\'\i sica de Cantabria (IFCA), CSIC-Universidad de
Cantabria, E-39005 Santander, Spain}
\author{Miguel A. Rodr\'\i guez}
\email{rodrigma@ifca.unican.es}
\affiliation{Instituto de F\'\i sica de Cantabria (IFCA), CSIC-Universidad de
Cantabria, E-39005 Santander, Spain}

\date{\today}

\begin{abstract}
In this work we perform a detailed study of the scaling properties of Lyapunov
vectors (LVs) for two different one-dimensional Hamiltonian lattices:
the Fermi-Pasta-Ulam and $\Phi^4$ models. In this case,
characteristic (also called covariant) LVs exhibit qualitative
similarities with those of dissipative lattices
but the scaling exponents are different and seemingly nonuniversal.
In contrast, backward LVs (obtained via
Gram-Schmidt orthonormalizations) present approximately 
the same scaling exponent in all cases, suggesting
it is an artificial exponent produced by the imposed orthogonality of these vectors.
We are able to compute characteristic LVs in large systems thanks to
a `bit reversible' algorithm, which completely
obviates computer memory limitations.
\end{abstract}

\pacs{05.45.Tp, 05.45.Jn, 05.45.Pq, 05.40.Jc}

\maketitle

\section{Introduction\label{sec:Intro}}

In dynamical systems
the sensitive dependence on initial conditions is readily quantified
by the Lyapunov exponents (LEs),
which measure the average growth rate of infinitesimal perturbations~\cite{ott}.
Spatially extended systems may exhibit spatiotemporal chaos (STC) and the
spectrum of LEs 
is a good indicator of the extensivity with the system
size~\cite{Livi86,Livi87}. 
The directions in tangent space associated with the LEs are
generically called Lyapunov vectors (LVs). These vectors convey important
dynamical information. For example, LVs have been useful
to discover and quantify the so-called hydrodynamic
modes~\cite{FPUHLM,LJHLM,morriss09},
to study extensivity properties~\cite{pomeau84,egolf00} and to address
predictability
questions in weather forecasting~\cite{kalnay,pazo10}, among other applications.

For many dissipative one-dimensional models~\cite{pik94,pik98} it is known that,
after a suitable logarithmic transformation, the infinitesimal perturbation
associated
with the largest LE, the main LV, belongs to the universality class of the
stochastic
Kardar-Parisi-Zhang (KPZ) equation~\cite{kpz} of surface growth.
The mentioned logarithmic transformation associates a ``surface'' with the LV,
leading to 
many interesting consequences; for instance: the scaling of finite-size corrections
and self averaging properties of the LEs.
The ``surface picture'' has demonstrated to be very powerful
and has been used to analyze 
finite perturbations \cite{lopez04,primo05,primo06} and singular vectors
\cite{pazo09} in STC.

The only homogeneous extended systems where the full correspondence 
between the main LV and the KPZ scaling
is known to break down are anharmonic Hamiltonian lattices.
In Ref.~\cite{pik01} it was determined, by numerical simulation of two different
oscillator lattice models,
that the main reason for the lack of KPZ scaling in Hamiltonian systems
can be traced back to the ubiquitous existence of long-range spatiotemporal
correlations in
the observables that control the LV dynamics. In Ref.~\cite{pik01} the authors
invoke
the KPZ equation with a long-range-correlated noise (instead of white noise) as
a minimal model
that accounts for their observations. 
Nevertheless, it remains unclear whether the KPZ equation with
spatio-temporal long-range correlated noise
is indeed the correct minimal model for the dynamics of the leading LV in
one-dimensional Hamiltonian lattices.
Further theoretical progress is needed to clarify this issue.

Recent studies~\cite{szendro07,pazo08} have extended the analysis
to LVs corresponding to the most unstable
directions (not only the leading one) in several dissipative systems. 
As reasoned above, it is clear that Hamiltonian lattices deserve a separate
study due to the peculiar behavior already observed for the main LV.

We employ the so called~\cite{legras96} \emph{characteristic} Lyapunov vectors
(CLVs) proposed many years ago
by Ruelle~\cite{ruelle79} because they reflect the bona-fide directions in
tangent space (see below).
CLVs have been recently employed to characterize several aspects of STC, such as
spatio-temporal correlations and extensivity~\cite{szendro07,pazo08,takeuchi09},
hyperbolicity~\cite{ginelli07,pavel09}, and
Oseledec splitting~\cite{yang08,yang09,kuptsov10}.
In addition CLVs have nice properties that may support
their use also for ensemble forecasting in atmospheric models~\cite{pazo10}.

The aim of this paper is to explore universality properties (if any) of CLVs
for Hamiltonian lattices, as these systems are already peculiar in what concerns
the main LV. In addition, we present an algorithm, specially designed
for Hamiltonian systems, to compute CLVs in large systems with modest
computer resources.

This paper is organized as follows: Sec.\ \ref{sec:Model} describes the employed
models and the relevant details of their numerical implementation. Sec.\
\ref{sec:SGP} gives the relevant details of the ``roughening surface''
picture, as well as some results concerning the temporal evolution of the
defined surface. In Sec.\ \ref{sec:SS} we investigate the spatial correlations
of the CLVs. The discussion of the obtained results
is made in Sec.\ \ref{sec:Disc}.

\section{the models and simulation details\label{sec:Model}}

\subsection{Phase-space dynamics}

The reference Hamiltonian for the one-dimensional coupled anharmonic lattice
models we are considering can be written as
\be
H=\sum_{i=1}^N\left[\frac{p_i^2}{2m_i} + V(q_{i+1}-q_i) + U(q_i)\right]
\label{GH},
\ee
where $N$ is the system size, and $V(x)$ and $U(x)$ are the nearest-neighbor
interaction and on-site potentials, respectively. The particles are assumed to
be
of unit mass $m_i=1$. The phase space coordinates (displacement and
momentum) are $\{q_i,p_i\}_{i=1}^N$; periodic boundary
conditions are assumed ($q_{_{N+1}}=q_{_1}$). In the following we shall consider
two models: (i) the Fermi-Pasta-Ulam (FPU) $\beta$ model \cite{FPUx},
characterized by $V(x) = x^2/2+x^4/4$ and $U(x)=0$,
and (ii) the $\Phi^4$ model~\cite{Pettini,Pettini2}, characterized by harmonic
interactions
$V(x)=x^2/2$ and by a double-well on-site potential $U(x)=-x^2/2+x^4/4$.

In our numerical simulations we have chosen 
as initial conditions the equilibrium value of the oscillators
displacements, i.e. $\mathbf{q}(0)\equiv\{q_i(0)=0\}_{i=1}^N$, and momenta
$\mathbf{p}(0)\equiv\{p_i(0)\}_{i=1}^N$ drawn from a Maxwell-Boltzmann
distribution at a temperature consistent with a given value of the energy
density $\epsilon\equiv E/N$. A value of $\epsilon=10$ has been chosen for the
FPU model,
since it is known that its dynamics is strongly chaotic
for $\epsilon\gg1$~\cite{Pettini,Pettini2}.
For the $\Phi^4$ model we chose $\epsilon=5$, as in Ref.~\cite{pik01}.

\subsection{Tangent-space dynamics}
\label{tsd}
To study the local dynamical stability of our system we introduce the
infinitesimal perturbations of the trajectory $\mathbf{\Gamma}\left(t\right)$
along all possible directions (position and momentum axes) of the phase space as
$\delta\mathbf{\Gamma}(t)\equiv\left(\delta\mathbf{q}\left(t\right),
\delta\mathbf{p}\left(t\right)\right)$, thus defining the $2N$-dimensional
\emph{tangent space}. These infinitesimal perturbations are governed by the
linear
equations
\begin{subequations}
\label{tan}
\begin{eqnarray}
\dot{\delta q_i}&=& \frac{\partial^2 H}{\partial q_i \partial p_i} \delta q_i
+ \frac{\partial^2 H}{\partial p_i^2} \delta p_i  \label{tan1}\\
\dot{\delta p_i}&=& -\frac{\partial^2 H}{\partial q_i^2} \delta q_i
- \frac{\partial^2 H}{\partial q_i \partial p_i} \delta p_i . \label{tan22}
\end{eqnarray}
\end{subequations} 
This linear evolution of infinitesimal perturbations implies the existence
of a linear operator (resolvent or propagator) ${\cal M}$ that links
perturbations
at different times: $\delta\mathbf{\Gamma}(t)={\cal
M}(t,t_0)\cdot\delta\mathbf{\Gamma}(t_0)$. 

According to Oseledec's multiplicative ergodic theorem~\cite{oseledec} the
remote past limit symmetric operator ${\Phi}_{\mathrm
b}(t)=\lim_{t_0\rightarrow-\infty}[{\cal M}(t,t_0)\cdot{\cal
M}^{*}(t,t_0)]^{1/[2(t-t_0)]}$ exists for almost any initial condition
$\mathbf{\Gamma}\left(t_0\right)$. The set of LEs is defined as
$\lambda_{\alpha}\equiv\ln\Lambda_\alpha$, where $\{\Lambda_\alpha\}$ are
the eigenvalues of $\Phi_{\mathrm b}(t)$. We label the LEs in decreasing order:
$\lambda_1 \ge \lambda_2 \ge \cdots \ge \lambda_{2N}$.
The standard procedure \cite{benettin80,shimada79} to
compute the ${\mathcal N}$ largest LEs resorts to periodic
Gram-Schmidt-orthonormalizations
of a set of ${\mathcal N}$ offset vectors evolved by Eqs.~(\ref{tan}).
The time-averaged values of the logarithms of the
normalization factors yield the
LEs $\left\{ \lambda_{\alpha}\right\}$. The set of vectors right after
each reorthonormalization $\{{\mathbf b}_\alpha(t)\}$ are the eigenvectors of
$\Phi_{\mathrm b}(t)$ \cite{ershov98}
and they are called \emph{backward} LVs (BLVs), following the nomenclature by
Legras and Vautard~\cite{legras96}.
Note that BLVs, apart from the main one ${\mathbf b}_1(t)$, are not
univocally defined because they depend on the scalar product for the
orthogonalization
(which also determines the adjoint operator ${\cal M}^{*}$;
${\cal M}^{*}={\cal M}^{\mathrm T}$ in Euclidean space). 

BLVs have the advantage of a straightforward calculation as they are simply
the byproduct of the standard method to compute LEs. However from the point of
view of the physical meaning there is another set of vectors, the already
mentioned CLVs
(also known as covariant LVs), which univocally
determine the direction in tangent space corresponding to each LE. The 
CLVs $\{{\mathbf g}_\alpha(t)\}$ were already defined by
Ruelle in 1979 \cite{ruelle79,eckmann}. These vectors are independent of 
the definition of the scalar product and readily signal the intrinsic stable and
unstable directions.
As a result CLVs are covariant with the linear
dynamics, ${\mathbf g}_\alpha(t) \propto {\cal M}(t,t_0) {\mathbf g}_\alpha
(t_0)$, wherewith it is automatically
guaranteed that the LEs are recovered in both, past and future, time limits:
\be
\lim_{|t| \to \infty} (t-t_0)^{-1} \ln ||{\cal M}(t,t_0) {\mathbf g}_\alpha
(t_0) ||= \lambda_\alpha . 
\ee

\subsection{Important numerical issues}

The evolution of the phase space trajectory
$\mathbf{\Gamma}(t)\equiv\left(\mathbf{q}\left(t\right),\mathbf{p}
\left(t\right)\right)$ is obtained integrating the $2N$ first-order
Hamilton equations of motion. We have used a 
symmetrical version of the velocity Verlet integrator
specially suited for long-time simulations~\cite{Tuck}, see Eq.~(\ref{sverlet})
in the Appendix.
The adopted time step value $\Delta t=0.01$ assures a faithful representation
of the Hamiltonian flow and a driftless average value of the total energy $E$
with a fluctuation level of $\Delta E/E\approx10^{-3}-10^{-4}$ depending on
the system size.

The computation of CLVs is not straightforward. We have used the method
proposed by Wolfe and Samelson in Ref.~\cite{wolfe07}, wherein all relevant
details are given. To find the ${\mathcal N}$th CLV one needs to compute:
(i) the first ${\mathcal N}$ BLVs and
(ii) a set of ${\mathcal N}-1$ vectors (forward LVs) integrating backwards in time the
perturbations that obey the adjoint operator of the linear dynamics (proceeding
as in the case of
BLVs, but using the {\em transpose} of the Jacobian matrix instead).

The problem of the time-reversed integration 
is that, although the employed algorithm to integrate the equations of motion
 derived from the Hamiltonian is explicitly time reversible,
the computed trajectories in phase space, obtained after
reversing all momenta, do not coincide with those traced by the time-forward
motion
(due to the effect of round-off errors and chaos sensitivity). We
have solved this problem by a suitable use of integer arithmetic operations,
which suppresses unwanted numerical effects. Thus the original
phase-space trajectory can be exactly traced back (see the Appendix for
technical details).
Note that our procedure consumes almost no computer memory because it only
requires to
store the set $\{{\mathbf b}_\alpha(t)\}_{\alpha=1}^{\mathcal N}$
at the times where the ${\mathcal N}$th CLV is going to be computed.
If the bit reversible
algorithm were not used the state of the system would have to be recorded
periodically to allow a faithful trajectory backtracking. Furthermore, if instead
of the Benettin method the QR method were used (as in Ref.~\cite{ginelli07}), 
the periodical storage of the {\bf R} matrices required by the latter would quickly
lead to a computer memory overflow.

\section{surface growth picture\label{sec:SGP}}

\begin{figure}
\includegraphics[width=0.90\linewidth,angle=0.0]{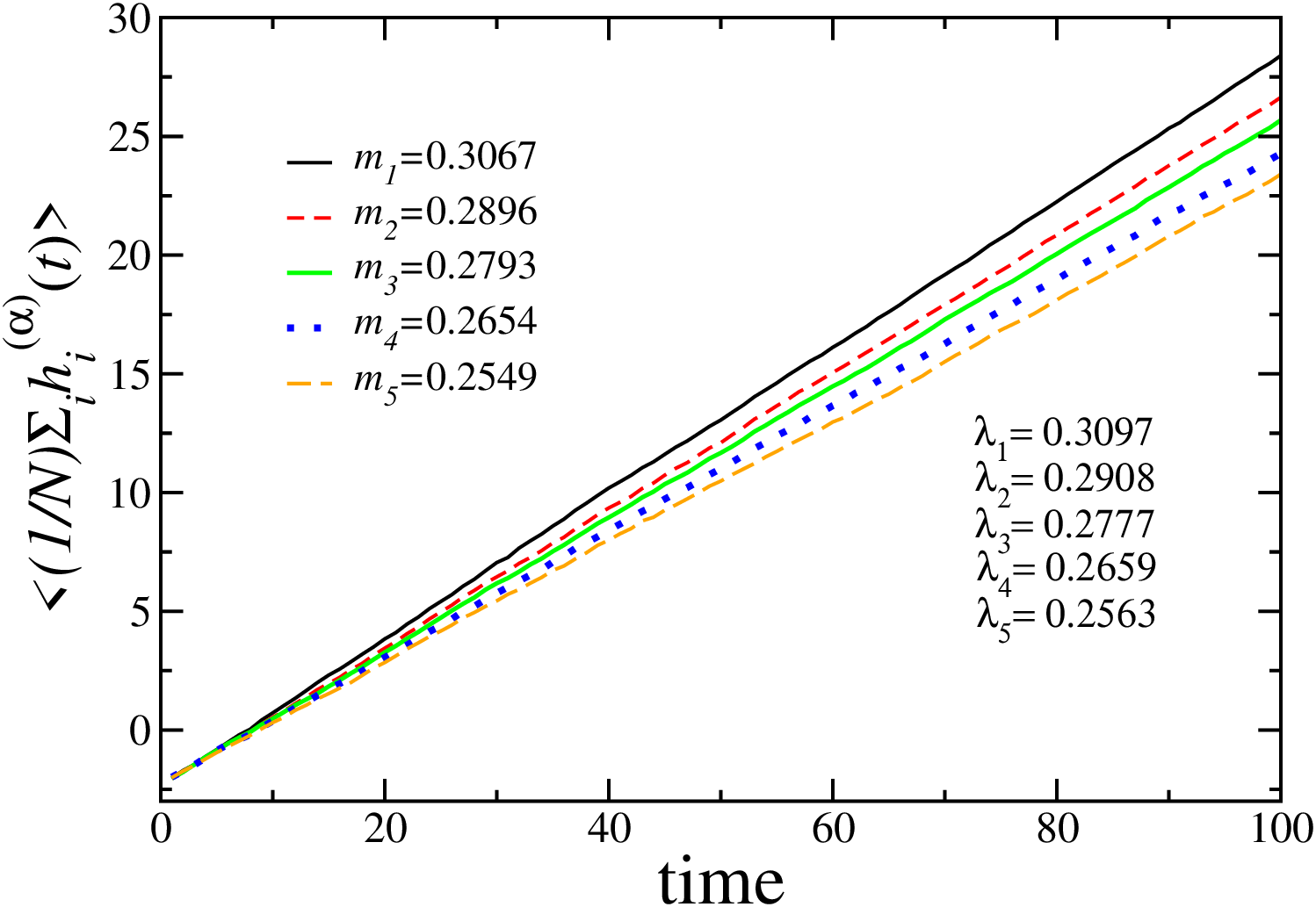}
\caption{(Color online) Evolution of the mean height of
surfaces $\langle\bar h_{\alpha}(t)\rangle$ associated with
small perturbations initialized along characteristic LVs with indices
$\alpha=1,\ldots,5$ from top to bottom (FPU model, $N=32$ and energy density
$\epsilon=10$). The curves are averages over 100 realizations.
The fitting slopes $m_{\alpha}$ are in good agreement with the
corresponding values of $\lambda_{\alpha}$.}
\label{fig:Lyap}
\end{figure}

For every CLV (likewise for BLVs) ${\bf
g}_{\alpha}(t)=\left(\delta\mathbf{q}^{\left(\alpha\right)}\left(t\right),
\delta\mathbf{p}^{\left(\alpha\right)}\left(t\right)\right)$ it is convenient to
define an associated ``surface''
\be
h^{(\alpha)}_i(t)=\ln\sqrt{[\delta{q}^{\left(\alpha\right)}_i\left(t\right)]^2+
[\delta{p}^{\left(\alpha\right)}_i\left(t\right)]^2}
\label{surface}
\ee
where, as before, index $i=1,\ldots,N$ plays the role of space. Hereafter
we refer to $\{h^{(\alpha)}_i\}$ as surfaces because
a relation, for $\alpha=1$,  between this kind of log-transformed LV and
stochastic surface growth
equations was discovered in Refs.~\cite{pik94,pik98}
(and proposed in Ref.~\cite{pik01} for Hamiltonian lattices).
Since $\bar h^{(\alpha)}(t) = (1/N)\sum_{i=1}^N
h^{(\alpha)}_i(t)$ is the logarithm of a norm,
the $\alpha$th LE corresponds to the average velocity of the corresponding
$\alpha$th surface, $\langle d \bar h^{(\alpha)}(t)
/dt\rangle=\lambda_{\alpha}$. Fig.~\ref{fig:Lyap} presents
a specific example of the nice properties of CLVs. Perturbations at $t=0$
along the first five CLVs are let to evolve freely, {\em i.e.}~obeying
Eqs.~(\ref{tan}),
and the mean height of the associated surfaces are computed versus time. The
average velocities
are fairly close to the corresponding LE values.

The main LV ${\mathbf g}_1={\mathbf b}_1$ in spatio-temporal
chaotic
systems is strongly localized in space~\cite{kaneko86,giaco91,falcioni91}, 
and transformation~(\ref{surface})
allows to unfold the spatial structure of the vector, which would be otherwise
hidden close to zero. The localization of the main LV is dynamic,
i.e.~there is a slow wandering of the localization region.
In Ref.~\cite{pik01} it was demonstrated by means of numerical simulations that, contrary
to dissipative systems and other systems with STC,
the surface associated with ${\mathbf g}_1$ does not fall into the universality
class of the KPZ equation.
This fact is attributed to the presence of long-range correlations in space and time
in Hamiltonian lattices~\cite{pik01}.

\begin{figure}
\includegraphics[width=0.80\linewidth,angle=0.0]{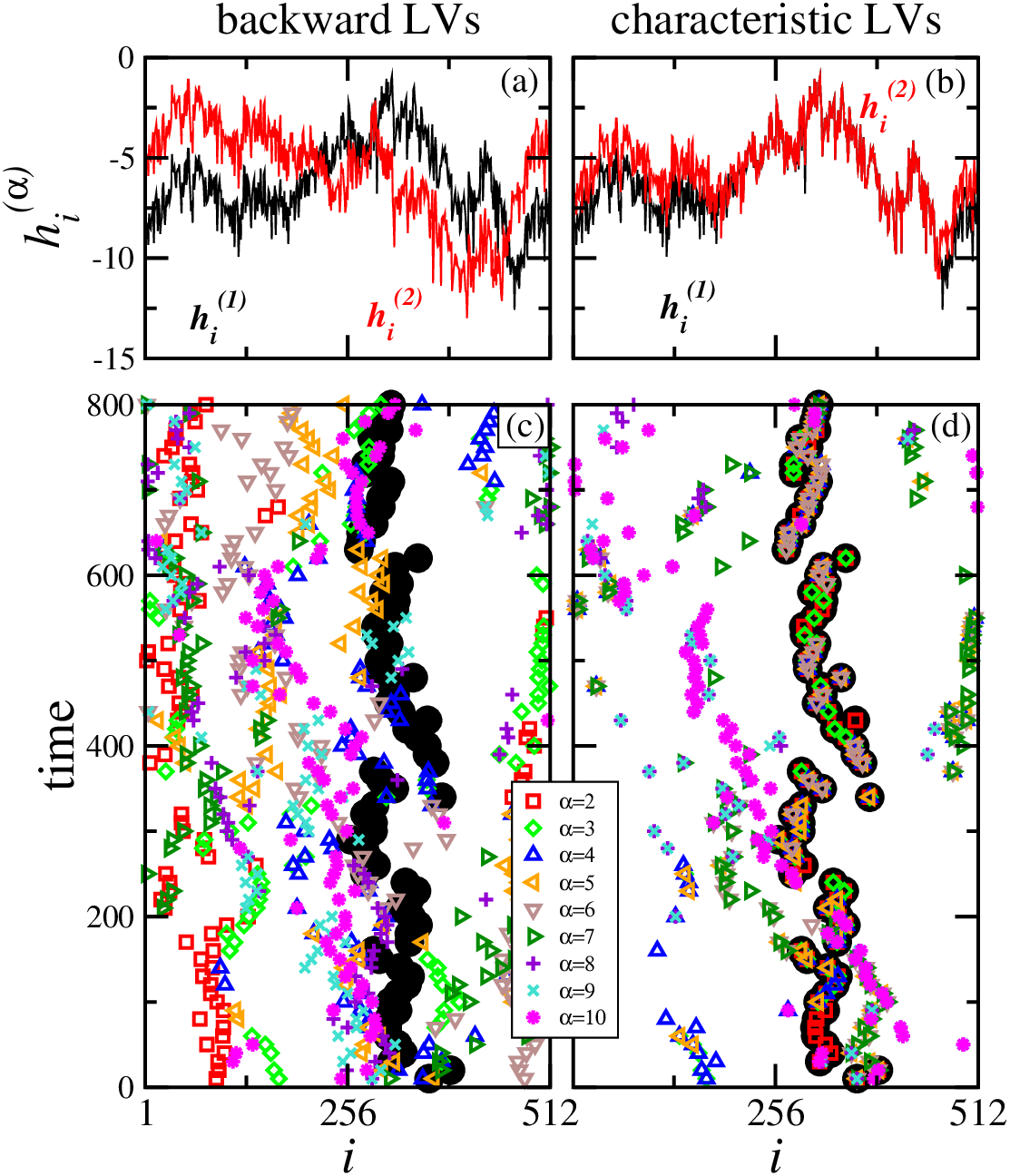}  
\caption{(Color online). (a,b) Snapshot of the LV-surfaces corresponding to
$\alpha=1,2$ for the FPU lattice with $N=512$. (c,d) Space-time evolution of the
localization sites for the first 10 LVs. The position of the localization site
for the main LV is indicated by filled circles. Other symbols correspond to
$\alpha=2,\ldots,10$.}
\label{fig:tevolfpu}
\end{figure}

For the FPU model we depict snapshots of
surfaces corresponding to the first and second LVs 
in Figs.~\ref{fig:tevolfpu} (a) for the BLVs and (b) for CLVs.
The vectors are strongly localized; notice that $h^{(\alpha)}_i$, Eq.~(\ref{surface}), is a logarithmic variable.
In Figs.~\ref{fig:tevolfpu}(c,d), we
plot the time evolution of the localization sites
corresponding to BLVs and CLVs, for $\alpha=1,\ldots,10$.
We define the localization site as the position $i$ where $h^{(\alpha)}$ takes its
largest value at a given time. 
It can be readily seen that the maxima corresponding to the BLV-surfaces are scattered all over the
spatial domain, just as in the case of the coupled-map-lattice (CML) studied in
Ref.~\cite{szendro07} where it was argued that this behavior is a byproduct of
the orthogonalization procedure and not a physical property of the
perturbation dynamics. 
On the contrary, CLVs present much more correlated localization sites.

\begin{figure}
\includegraphics[width=0.95\linewidth,angle=0.0]{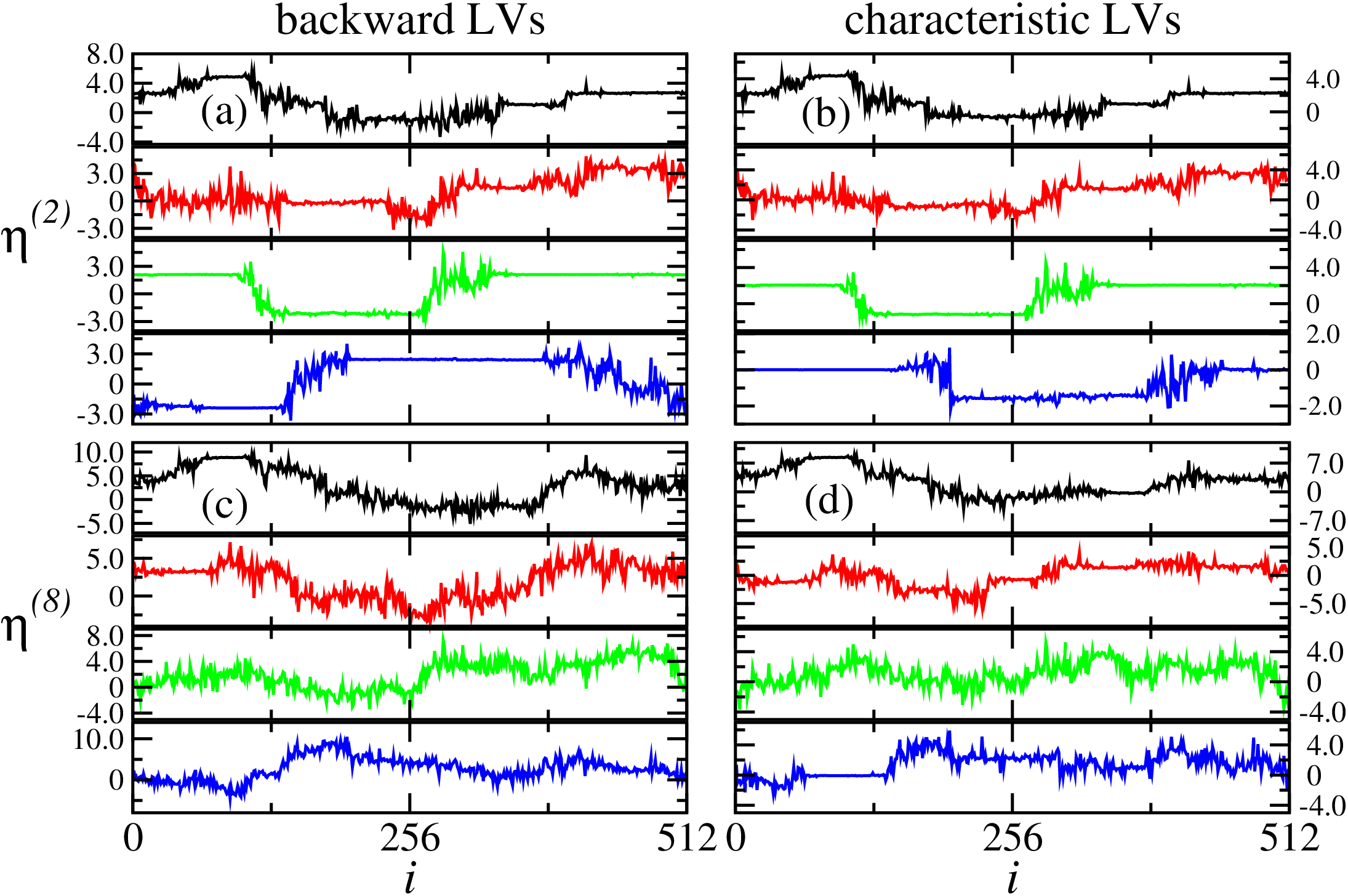}
\caption{(Color online) Snapshots of the field $\eta_i^{(2)}$ at four different
times for (a) backward LVs and (b) characteristic LVs.
(c,d) Same as (a,b) but for the field $\eta_i^{(8)}$.
Note that the typical length of the plateaus is smaller in this case.
(FPU model with $N=512$ and $\epsilon=10$.)}
\label{fig:eta}
\end{figure}

The scaling properties of LVs corresponding to LEs smaller than the first one
have been recently reported for spatio-temporally chaotic dissipative 
systems~\cite{szendro07,pazo08}.
These works revealed that LV-surfaces are piecewise copies of the main one.
This is readily seen defining the difference-field
\begin{equation}
\eta_i^{(\alpha)}\equiv h_i^{(\alpha)}-h_i^{(1)}.
\end{equation}
We show in Fig.~\ref{fig:eta}, using CLVs, that this qualitative feature
holds for Hamiltonian lattices as well (it is irrelevant that the first LV
does not belong to the KPZ universality class).
As in dissipative systems~\cite{szendro07,pazo08},
the typical plateau size of $\eta^{(\alpha)}$ decreases as $\alpha$ grows,
and beyond some $\alpha_{\mathrm{max}}$
this simple picture does not hold: Figure~\ref{fig:eta}(b) shows that for $\alpha=8$
the plateaus are smaller than for $\alpha=2$, Fig.~\ref{fig:eta}(a).

\section{spatial structure\label{sec:SS}}

In this section we perform a quantitative description of the spatial
correlations of the LV-surfaces $h_i^{(\alpha)}(t)$. We compute the stationary
structure factor
$S_{\alpha}(k)=\lim_{t\rightarrow\infty}\langle\hat{h}^{(\alpha)}(k,t)\hat{h}^{
(\alpha)}(-k,t)\rangle$, where
$\hat{h}^{(\alpha)}(k,t)=N^{-1/2}\sum_j\exp(2\pi\imath \, k  \,j)
h_j^{(\alpha)}(t)$, with
$\langle\ldots\rangle$ indicating an average over different system trajectories
(which correspond to different random initial conditions).

\begin{figure}
\includegraphics[width=0.9\linewidth,angle=0.0]{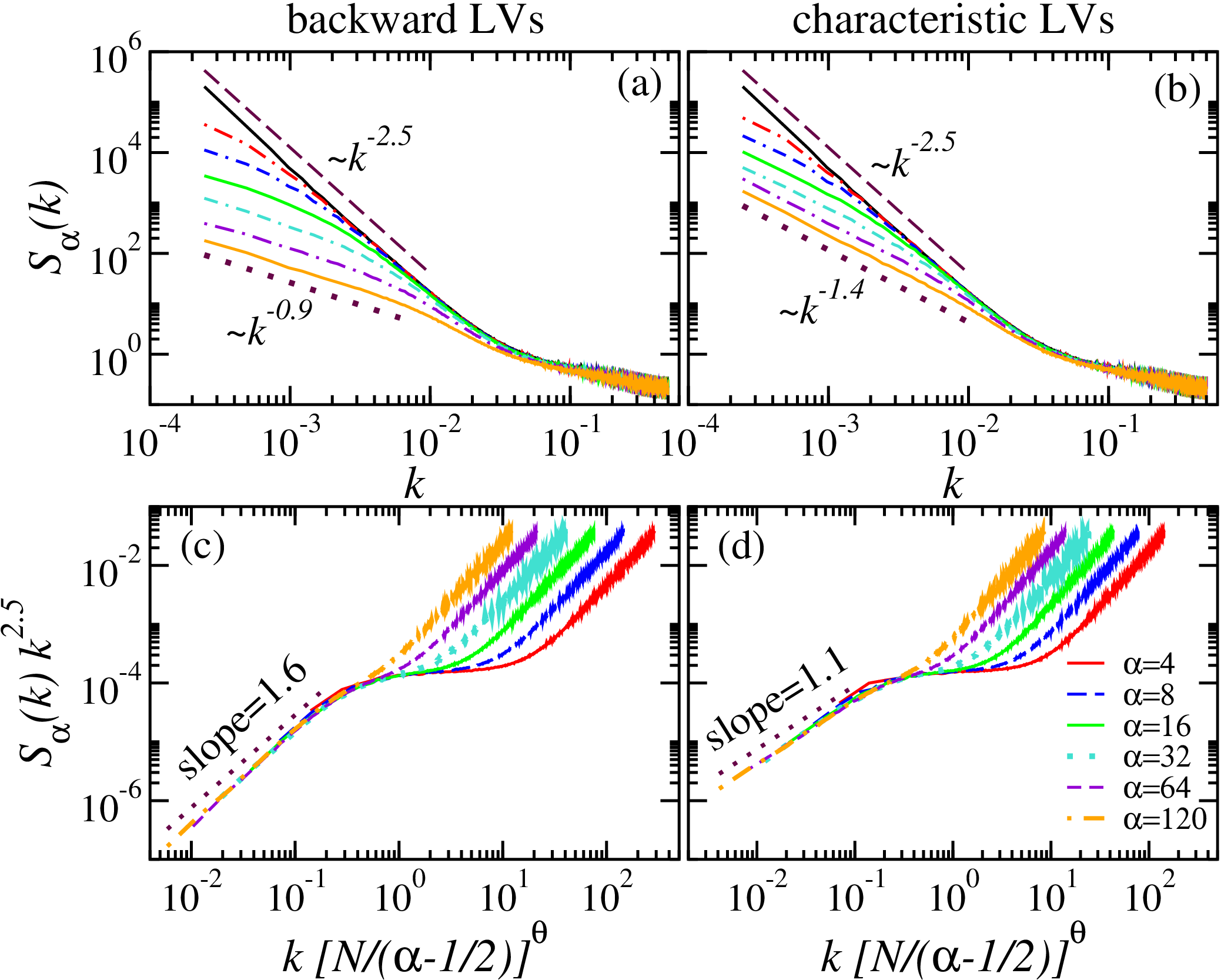}
\caption{(Color online) Structure factors $S_\alpha(k)$ of the FPU model with $N=4096$,
for (a) backward and (b) characteristic LV-surfaces, $\alpha=1,4,8,16,32,64,120$ from top to bottom.
(c,d) Data collapse through scaling relation~(\ref{SScaling})
with $\gamma=-2.5$ and fitting parameters $\theta=0.9$ (c) and $\theta=0.8$ (d); index $\alpha$
increases from right to left.}
\label{fig:hbcfpu}
\end{figure}

\begin{figure}
\includegraphics[width=0.9\linewidth,angle=0.0]{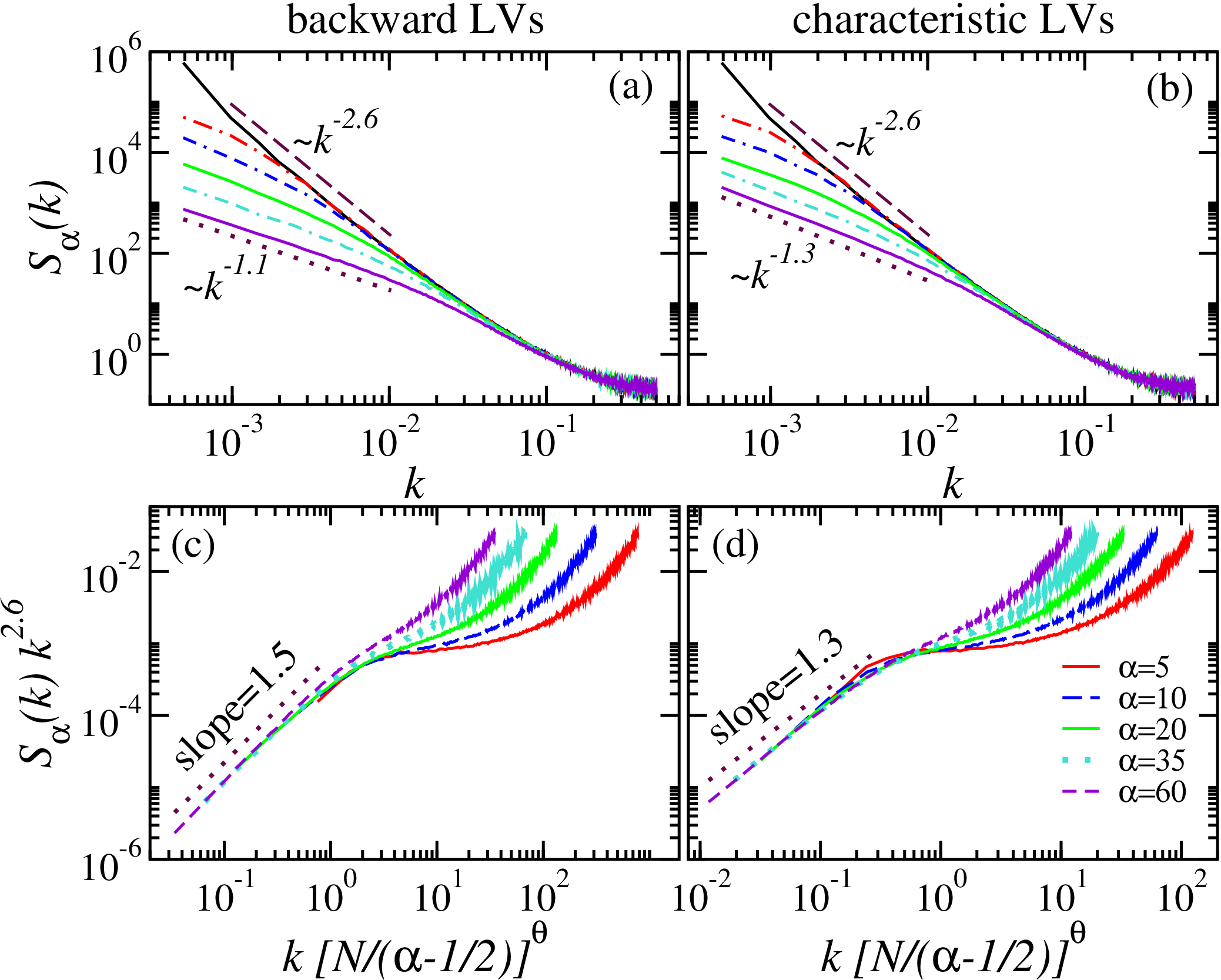}
\caption{(Color online) Structure factors $S_\alpha(k)$ of the $\Phi^4$ model with $N=2048$,
for (a) backward and (b) characteristic LV-surfaces, $\alpha=1,5,10,20,35,60$ from top to bottom.
(c,d) Data collapse through scaling relation~(\ref{SScaling})
with $\gamma=-2.6$ and fitting parameters $\theta=1.2$ (c) and $\theta=1.0$ (d); index $\alpha$
increases from right to left.}
\label{fig:hbcphi4}
\end{figure}

Figures~\ref{fig:hbcfpu}(a,b) and~\ref{fig:hbcphi4}(a,b) show the structure
factors
of a representative set of LV-surfaces for FPU and $\Phi^4$ models,
respectively.
For the main LV surface, $\alpha=1$,
the short wavenumber scaling exponent $\gamma$ of the structure factor
($S(k)\sim k^{\gamma}$)
is clearly different from the one expected for the KPZ
universality class in one dimension ($\gamma=-2$)
and the obtained values, $\gamma=-2.5$ (FPU) and $\gamma=-2.6$ ($\Phi^4$), are
consistent with 
those reported by Pikovsky and Politi~\cite{pik01}.

As explained in Sec.~\ref{tsd}, for $\alpha > 1$ one must distinguish
between backward and characteristic LVs. Figures~\ref{fig:hbcfpu}(a,b) 
and~\ref{fig:hbcphi4}(a,b) 
evidence that both vector types indeed have different spatial structures.
The structure factors of BLV-surfaces asymptotically decay 
with exponents $\gamma=-0.9$ (FPU) and
$\gamma=-1.1$ ($\Phi^4$), which are close to the value $\gamma=-1$
reported for dissipative systems in Refs.~\cite{szendro07,pazo08}.
These results suggest that the value $\gamma\approx-1$ for BLVs in our
Hamiltonian systems
has a geometric origin and is related to the Gram-Schmidt orthonormalization.
However, CLVs display exponents $\gamma=-1.4$ (FPU) and
$\gamma=-1.3$ ($\Phi^4$) which, at least for the FPU model,
are different from the values $-1.2$ or $-1.15$
reported in dissipative systems~\cite{szendro07,pazo08}.

The fact that LVs corresponding to the most expanding directions
are (in the surface representation) piecewise copies of the main LV
translates into the existence of crossover wavenumbers where the
structure factors bend. Each structure factor $S_\alpha$ presents a knee 
at a certain wavenumber $k_\alpha^\times$ that is related to the typical 
plateau length of the difference field $\eta^{(\alpha)}$. 
For both, FPU and $\Phi^4$, models $k_\alpha^\times$ scales with $\alpha$
approximately as $k_\alpha^\times\sim \left[(\alpha-1/2)/N\right]^\theta$,
with $\theta \approx 1$ (as in Ref.~\cite{szendro07} for a dissipative CML).
We choose to use $\alpha-1/2$ instead of $\alpha$, as it is customary when plotting
the Lyapunov spectrum (this is actually irrelevant in the thermodynamic limit $N\to\infty$).
This is verified in Figs.~\ref{fig:hbcfpu}(c,d) and \ref{fig:hbcphi4}(c,d)
through a data collapse of the structure factors via the scaling relation
\be
S_{\alpha}(k)k^{-\gamma}=g(k/k_\alpha^\times) \label{SScaling},
\ee
where $g(u)=$const for $u\gg1$ and $g(u)\sim u^{\sigma}$ for $u \ll 1$.

\section{discussion and conclusions\label{sec:Disc}}

In a recent paper~\cite{pazo08} we established (as suggested by~\cite{pik98})
a minimal stochastic model for LVs $w(x,t)$ in spatiotemporal chaos:
\begin{equation}
\partial_{t} w(x,t)= \xi(x,t) w(x,t) + \partial_{xx} w(x,t),
\label{minimal1}
\end{equation}
where $\xi(x,t)$ is a white noise term that accounts for the chaotic fluctuations.
The generic asymptotic solution of Eq.~(\ref{minimal1}) models the main LV, whereas
``saddle solutions'' (see \cite{pazo08} for details) correspond to CLVs
with $2\le \alpha \le \alpha_{\mathrm{max}}$. This correspondence between a
(linear) stochastic equation and LVs is
supported by the existence of common scaling exponents~\cite{pazo08}. 
The generic solution of~(\ref{minimal1}) has the same sign
in all the domain, and this allows to transform Eq.~(\ref{minimal1})
into the KPZ equation through the Hopf-Cole transformation
$h(x,t)=\ln|w(x,t)|$. In contrast, the saddle-solutions of~(\ref{minimal1})
vanish at several points, what precludes KPZ as a valid equation for CLVs
other than for $\alpha=1$. 

In Hamiltonian systems, the anomalous scaling exponent ($\gamma\ne 2$)
of the main LV was traced back~\cite{pik01} to the
long-range correlations of the multipliers driving the linear equations~(\ref{tan}).
For Hamiltonians of the form~(\ref{GH}),
these multipliers are functions of the displacements $\{q_i\}$ with specific properties for
FPU and $\Phi^4$ models.
The minimal model proposed in~\cite{pik01} for the main LV-surface was the
KPZ equation [or~(\ref{minimal1}) reversing the Hopf-Cole transformation] with long-range
correlated noise. Unfortunately, theoretical expressions for  $\gamma$~\cite{medina89,Barabasi}
consider the KPZ equation with either spatially or temporally-correlated noise (but not both).
The exponent $\gamma$ may take the same value with different combinations of
spatial and temporal long-range correlations. Hence there is not a univocal relation between
correlations and $\gamma$.

Minimal models are important as they are more amenable to theoretical analysis,
which should allow to distinguish different universality classes\footnote{For instance the
heat conductivity is anomalous for the FPU chain, whereas it is normal for the $\Phi^4$
lattice~\cite{heatfpuphi4}.}.
Our present work has the value of giving more constraints to
the minimal stochastic model for perturbations in Hamiltonian lattices. 
A minimal model should reproduce the scaling
properties of both the main LV and sub-dominant LVs. These sub-dominant ($\alpha\ge 2$) LVs
have been the subject of the present study.
A possible alternative to~(\ref{minimal1}) pointed out in~\cite{pik98} is 
the time-reversible equation:
\begin{equation}
\partial_{tt} w(x,t)= \xi(x,t) w(x,t) + \partial_{xx} w(x,t).
\label{minimal2}
\end{equation}
Future work is need to find out the true minimal model for LVs in Hamiltonian lattices.
In any case, our results should be a guidance in the search of such a minimal model
for Hamiltonian systems, since any suitable minimal model must produce surfaces with the scaling properties in
Figs.~\ref{fig:hbcfpu} and~\ref{fig:hbcphi4}.

Finally, a bit reversible algorithm, which operates with integer arithmetic, has been implemented to take advantage of the time reversibility of Hamiltonian systems. Although trajectory reversibility is not strictly required to compute the CLVs (in previous works~\cite{szendro07,pazo08} CLVs have been computed in
non-reversible dissipative systems), our methodology makes use of the aforementioned reversibility to completely bypass the need to store the phase-space trajectory in order to maximize the efficiency of the computation of the CLVs from the intersection of backward and forward LV subspaces if only limited computer resources are available.

\acknowledgments
M.R.B. acknowledges financial support from PROMEP and CONACyT, M\'exico, as
well as the warm hospitality of the U.A.E.M. during the first stages of this
work.
D.P.~acknowledges support by CSIC under the Junta de Ampliaci\'on de
Estudios Programme (JAE-Doc).
Financial support from the Ministerio de Ciencia e Innovaci\'on (Spain) under
projects
FIS2009-12964-C05-05 and CGL2007-64387/CLI is acknowledged.

\appendix*

\section{Bit reversible algorithm}

Since the so called `bit reversible' technique has been so far implemented using
the standard Verlet integrator (which does not considers the momenta explicitly)
and employed mainly for studies of time reversibility in Lennard-Jones
fluids~\cite{Verlet,Romero,Komatsu}, (although it has also been applied in
cases in which the so-called smoothed-particle continuum mechanics becomes
isomorphic to molecular dynamics, see Ref.~\cite{Hoover}) we will give a brief
explanation of its current implementation in order to make this paper
self-contained. Starting from the initial condition ${\mathbf\Gamma}(0)$, the
phase-space point ${\mathbf\Gamma}(\Delta t)$ is obtained by means of the
symmetrical velocity Verlet integrator~\cite{Tuck} written, in floating-point
arithmetic, as
\begin{subequations}
\label{sverlet}
\begin{eqnarray}
p_i\left(\frac{\Delta t}{2}\right)&=&p_i(0) + \frac{\Delta t}{2}F_i[q_i(0)], 
\label{sverlet1}\\
q_i(\Delta t)&=&q_i(0) + \frac{\Delta t}{m_i}p_i\left(\frac{\Delta t}{2}\right),
\label{sverlet2}\\
p_i(\Delta t)&=&p_i\left(\frac{\Delta t}{2}\right) + \frac{\Delta
t}{2}F_i[q_i(\Delta t)], \label{sverlet3}
\end{eqnarray}
\end{subequations} 
where $F_i$ is the total force on the $i$th oscillator. The bit reversible
version of algorithm~(\ref{sverlet}) employs an integer representation of phase
space instead of the conventional continuous phase space. To accomplish such
transformation, for the considered lattices
the minimum distance by which the
phase space is discretized is defined as $\Delta L=N/2^n$, where $N$ is the
system size and $2^n$ is the largest integer value if $n$-bit integers are
employed. Because of the discretization, the phase space coordinates are
represented by integers, i.e. $\{\mathbf{iq},\mathbf{ip}\}$. Therefore the
evolution equations can be recast in the following form:
\begin{subequations}
\label{sverleti}
\ba
\mathrm{ip}_i\left(\frac{\Delta t}{2}\right)&=&\mathrm{ip}_i(0) +
\sum_{j=i-1}^{i+1}\left\{\frac{\Delta
t}{2}F_{ij}[q_i(0)]\right\}_{\mathrm{Integer}}, \label{svi1} \\
\mathrm{iq}_i(\Delta t)&=&\mathrm{iq}_i(0) + \left\{\frac{\Delta t}{
m_i}p_i\left(\frac{\Delta t}{2}\right)\right\}_{\mathrm{Integer}},\label{svi2} \\
\mathrm{ip}_i(\Delta t)&=&\mathrm{ip}_i\left(\frac{\Delta t}{2}\right) +
\sum_{j=i-1}^{i+1}\left\{\frac{\Delta t}{2}F_{ij}[q_i(\Delta
t)]\right\}_{\mathrm{Integer}}, \label{svi3}
\ea
\end{subequations}
where $F_{ij}$ is a partial force from the $j$-th nearest neighbor on the $i$-th
oscillator. It should be noted that the discrete coordinates
$\{\mathbf{iq}(t),\mathbf{ip}(t)\}$ are integers, and the actual phase-space
coordinates are obtained as $\mathbf{q}(t)=\mathbf{iq}(t)\Delta L$ and
$\mathbf{p}(t)=\mathbf{ip}(t)\Delta L$. The second terms in the right hand side
of Eqs.~(\ref{sverleti}) are calculated based on the continuous phase space
variables $\{\mathbf{q},\mathbf{p}\}$ and the values in brackets
$\{\cdot\}_{\mathrm{Integer}}$ are converted to integers; in this way the total
momentum is exactly zero at all times during the simulation~\cite{Verlet}.

With the aforementioned implementation time reversibility is achieved exactly
throughout the simulations performed, which were rather long. To give an example, for the
$\Phi^4$ model a simulation of $5\times10^8$ time steps, after a transient of $1.5\times10^8$,
was needed to obtain the reported results. The situation is definitely better for
the FPU model, where only $5\times10^6$ time steps, with a transient of $5\times10^5$,
were sufficient for the employed system size.
Due to the conversion process to integer, energy is not
exactly conserved. Nevertheless, the exact-time-reversibility of the integration
algorithm precludes any systematic drift in the total energy. We have confirmed
that the fluctuation of the total energy by the bit reversible simulations is
equivalent to that by conventional floating-point simulations.

\bibliographystyle{prsty}

\end{document}